\documentclass[preprintnumbers,amsmath,amssymb,showpacs, twocolumn]{revtex4}

\usepackage{graphicx}
\usepackage{dcolumn}
\usepackage{bm}

\begin{document}

\title{Quantum secret sharing protocol between multiparty and multiparty  with single
photons and unitary transformations}

\author{ YAN Feng-Li$^{1}$, GAO Ting$^{2}$, LI You-Cheng$^{1}$ }

\affiliation {$^1$ College of Physics  Science and Information
Engineering, Hebei Normal
University, Shijiazhuang 050016, China\\
$^2$ College of Mathematics and Information Science, Hebei Normal
University, Shijiazhuang 050016, China }

\date{\today}

\begin{abstract}
We propose a  scheme of quantum secret sharing between Alices' group
and Bobs' group  with single photons and unitary transformations. In
the protocol,
 one member in Alices' group prepares a sequence of single photons in one of four different
states, while other members   directly encode their information on
the sequence of single photons via unitary operations, after that
the last member  sends the sequence of single photons to Bobs'
group. Then Bobs except for the last one do work similarly. Finally
last member in Bobs' group measures the qubits. If Alices
 and Bobs guarnnted the security of  quantum channel by some tests,
 then the qubit states  sent by last member of Alices' group  can be used as key bits for secret sharing.
It is shown that this
scheme is safe. \\
\end{abstract}


\maketitle

Quantum cryptography or quantum key distribution  is a remarkable
application of quantum information. It has attracted widespread
attention since the seminal work on quantum key distribution by
Bennett and Brassard \cite {BB84} and Ekert  \cite {E91}. So far
there are many quantum key distribution theoretical protocols  \cite
{BB84,E91,B92,BBM92,GV95,HIGM95,KI97,B98,HKH98,CaPRL00,CaPRA00,LLPra02,XLGPra02,
PBTB95,SongPra04,WangPrl04,WangPr07,GRTZrmp02}. Some of them were
demonstrated in the laboratory and will be used in the future secret
communications. With quantum mechanics, other cryptographic tasks
can be realized, such as quantum secure direct communication \cite
{LongGuiLuzongshu,
SIpra99,SIpra00,Beige02,Wprl03,Cprl03,BFprl02,DLLpra03,DLpra04,WDLLLpra,CLcpl,
CLpra04,ZMLpla04,ZMLpla204,ZXFZpra04,CScpl04, YZepjb04, GYWjpa05}
 and quantum
secret sharing (QSS) \cite
{HBB99,XLDPpra04,KKIpra99,TOIpra05,LSBSLprl04,DengPra05,DLLZWpra05,TZGpra01,
GottesmanPra00,NQIpra01,GLprl99,KBBpra02,DLZpla05,DLZZpra05E,ZLMpra05,ZMpra05,ZYMLpra05,
DLZZpra05,DZLpla05,GGpla03,YGpra05}.

Assume that there are  a manager  Alice, and her   agents Bob and
Charlie, who are at remote places.  One of the agents, Bob or
Charlie, is not entirely trusted by Alice, and unfortunately she
does not know who is the honest one. However Alice  knows that if
Bob and Charlie cooperate, the honest one will keep the dishonest
one from doing anything wrong. By means of the secret sharing Alice
can instruct Bob and Charlie to complete a task safely.
  The idea of
secret sharing  is that a secret of  Alice, is shared between her
agents,  Bob and Charlie,  in such a way that it can only be
reconstructed if both collaborate.  However, as a matter of fact,
there is not any classical mean to establish a secret sharing
between Alice and two distant  agents Bob and Charlie. Amazingly,
the principle of quantum mechanics has now provided the foundation
stone to QSS referring to the implementation of the secret sharing
task. The first QSS protocol of quantum secret sharing \cite {HBB99}
 was put forward by Hillery, Bu$\rm{\check{z}}$ek and Berthiaume using entangled
three-photon Greenberger-Horne-Zeilinger (GHZ) states. Xiao et al
\cite {XLDPpra04} reformulated the protocol \cite {HBB99} into
arbitrary number party case.  Karlsson et al \cite {KKIpra99}
realized secret sharing as in Ref. \cite {HBB99} with two-particle
quantum entanglement. They also discussed how to detect
eavesdropping, or how to detect a dishonest party in the protocols.
Furthermore, the concept of  quantum sharing of classical secrets
has also been generalized to the sharing of secret quantum
information, which is often called quantum state sharing \cite
{DengPra05,DLLZWpra05}. In a more general case, notably for secure
key management, a $t$-$out$-$of$-$n$ protocol (or $(t,n)$-threshold
scheme) with $1\leq t\leq n$ spreads a secret to $n$ participants in
a way that any $t$ participants can reconstruct it \cite {TOIpra05}.
Recently, Lance et al reported an experimental demonstration of a
(2,3) threshold QSS protocol \cite {LSBSLprl04}.

Most of the existing QSS protocols use  entangled states \cite
{HBB99,XLDPpra04,KKIpra99,TZGpra01,GottesmanPra00,NQIpra01,GLprl99,KBBpra02,DLZpla05,ZMpra05,ZYMLpra05}.
 However  it is believed  that  implementing such  multiparty secret sharing  tasks is not always so easy,
 as the efficiency of preparing  entangled states is very low \cite {BPDWZprl99,PDGWZprl01},
 at the same time the efficiency of the existing QSS
 protocols using quantum entanglement  can only approach
 50\%. Recently, a scheme for quantum secret sharing without entanglement has been proposed
 by Guo and Guo \cite {GGpla03}. They presented an idea to directly encode the qubit of quantum
 key distribution and accomplish one splitting a message into many parts to achieve
 one-to-multiparty secret sharing  only by product states. The theoretical efficiency is doubled to approach
 100\%. Deng et al also presented a one-to-multiparty secret sharing by using single photons \cite {DZLpla05}.
 Zhang et al proposed an QSS protocol by using single
 photons and unitary transformations \cite {ZLMpra05,DLZZpra05,DLZZpra05E}.

 Signatures on documents, authentications,
 encryptions, and decryptions are often needed by   more than one
 person, especially by all persons of two groups in a trade. Therefore, it is needed
 to share a
 secret between many parties and many parties. That is, it is  required the secret message shared by two groups,
 in such a way that neither any subset of each group nor the union of a subset of group 1 and  a subset of group 2 can
 extract the secret message, but the entire group 1 or group 2 can. More recently, we proposed a quantum secret sharing
 scheme employing two conjugate bases, i.e. four states, of single
 qubits to achieve the secret sharing between multiparty and
 multiparty with a sequence of single photons \cite {YGpra05}.  The weakness of our protocol was pointed out by Deng et al
 \cite {DYLLZGquph05} and Li et al \cite {LCHpra}, they also gave a little
 modification to avoid the flaw. We have avoided the flaw pointed out in Ref. \cite {LCHpra} with a little improvement of
  our protocol \cite {YGLScienceinChina}. Another QSS protocol between
  multiparty  and multiparty using three rather than two conjugate
  bases was suggested by us  \cite {GYquant062}.

In present Letter, we will combine the protocol of Zhang's in Ref.
\cite {ZLMpra05} and ours \cite {YGpra05, YGLScienceinChina} to
figure out a scheme of quantum secret sharing between multiparty and
multiparty with single photons and unitary transformations. The
security of this  protocol is discussed.

 Let us suppose  that group 1 and group 2 are located in different places, and there
 are members Alice 1, Alice 2, $\cdots$, Alice $m$, and  Bob 1,
 Bob 2, $\cdots$, Bob $n$  in group 1, and group 2 respectively. Here  $m\geq 2$ and $n\geq 2$. Group 1
 and group 2 want quantum secret sharing such that neither part of each group nor the combination
of a part of group 1 and a part of group 2 knows the key, but only
all members of each group can collaborate to determine what the
string (key) is. A protocol employing two conjugate bases  going as
follows will meet the  goal of quantum secret sharing.

M1. Alice 1 generates two random classical bit strings $A_1=\{a^1_1,
a^1_2, \cdots, a^1_{N}\}$ and $B_1=\{b^1_1, b^1_2, \cdots,
b^1_{N}\}$, where $a^1_k$ and $b^1_k$ are uniformly taken from $\{0,
1\}$. Based on these two strings $A_1$ and $B_1$ she then makes  a
block of $N$ qubits (single photons),
\begin{eqnarray}\label{Alice1}
   |\Psi^1\rangle & = &\otimes_{k=1}^{N}|\psi_{a^1_kb^1_k}\rangle,
     \end{eqnarray}
where $a^1_k$ is the $k$th bit of $A_1$ (and similar for $B_1$),
each qubit $|\psi_{a^1_kb^1_k}\rangle$ is in one of the four states
\begin{eqnarray}\label{singlequbits}
  |\psi_{00}\rangle &=& |0\rangle,\label{singlequbit1}\\
  |\psi_{10}\rangle &=& |1\rangle,\\
  |\psi_{01}\rangle &=& |+\rangle=\frac{|0\rangle+|1\rangle}{\sqrt{2}},\\
  |\psi_{11}\rangle &=& |-\rangle=\frac{|0\rangle-|1\rangle}{\sqrt{2}}.\label{singlequbit6}
\end{eqnarray}
The basis is determined by the  value of $b^1_k$. If $b^1_k$ is 0
then $a^1_k$ is encoded in the $Z$ basis $|0\rangle, |1\rangle$; if
$b^1_k$ is 1 then $a^1_k$ is encoded in the $X$ basis
$|+\rangle=\frac{|0\rangle+|1\rangle}{\sqrt{2}}$,
$|-\rangle=\frac{|0\rangle-|1\rangle}{\sqrt{2}}$. As the four states
are not all mutually orthogonal, so there is no measurement which
can distinguish between  all of them with certainty. Alice 1 then
delivers $|\Psi^1\rangle$ to Alice 2 over their public quantum
communication channel.

M2. When Alice $i$ $(i=2,3,\cdots, m)$ receives $N$ signals
delivered by Alice $i-1$ she use a special filter to prevent the
invisible photons from entering the operation system \cite
{CaiQY06,DLZZpra05E}. Then she chooses randomly a sufficiently
enough subset of photons as the samples for eavesdropping check.
First, let sample photon signals go to  a photon number splitter
(PNS: 50/50), and then measures each signal in the measurement basis
(MB) $Z$, or $X$ choosing randomly \cite {DYLLZGquph05}. Evidently
if more than two photons in one signal are measured, then Alice $i$
will abort the communication. Moreover, she  analyzes the error rate
$\varepsilon_s$ of the samples by means that she requires Alice 1,
Alice 2, $\ldots$, Alice $i-1$ to tell her the original states of
the samples and the operations implemented by them. If the error
rate is reasonably lower than the threshold $\varepsilon_t$, Alice
$i$ goes ahead,  otherwise she aborts it.

M3.  Alice $i$  $(i=2, 3, \cdots, m)$ generates a quaternary  string
$A_i=\{a^i_1, a^i_2, \cdots, a^i_{N}\}$ and a binary string
$B_i=\{b^i_1, b^i_2, \cdots, b^i_{N}\}$, where $a^i_k$ and $b^i_k$
are uniformly chosen from $\{0, 1, 2, 3\}$ and $\{0, 1\}$,
respectively.
    For each $|\psi_{a^{i-1}_k b^{i-1}_k}\rangle$ of the $N$ qubit state
 $|\Psi^{i-1}\rangle=\otimes_{i=1}^{N}|\psi_{a^{i-1}_k b^{i-1}_k}\rangle,$  she implements  the
    operation $\sigma_0$, $\sigma_1$, $\sigma_2$ or $\sigma_3$
 on it depending on the corresponding   $a^i_k$ of $A_i$ is 0, 1, 2
or 3, respectively.
 At the same time she  operates the qubit with $I$ or a
Hadamard operator $H$ according to the bit $b^i_k$ in $B_i$ is 0 or
1, respectively. Here
\begin{eqnarray} &&\sigma_0=I=|0\rangle\langle 0|+|1\rangle\langle 1|,\nonumber\\
&&\sigma_1=\texttt{i}\sigma_y=-|1\rangle\langle 0|+|0\rangle\langle 1|,\nonumber\\
&&\sigma_2=\sigma_z=|0\rangle\langle 0|-|1\rangle\langle 1|,\nonumber\\
&&\sigma_3=\sigma_x=|0\rangle\langle 1|+|1\rangle\langle 0|,\nonumber\\
&&
H=\frac{1}{\sqrt{2}}(|0\rangle+|1\rangle)\langle0|+\frac{1}{\sqrt{2}}(|0\rangle-|1\rangle)\langle1|.
\end{eqnarray}
These unitary operations made by  Alice $i$ are  equal to the
encryption on the states of  single photons.
 The resulting state of this qubit is denoted by $|\psi_{a^i_k
b^i_k}\rangle$.

M4. Alice $i$ sends the photons ($N$ qubit state
$|\Psi^i\rangle=\otimes_{k=1}^{N}|\psi_{a^i_k b^i_k}\rangle$) to
Alice $i+1$ ($i=2, 3, \cdots, m-1$). Alice $m$ sends  $N$-qubit
state $|\Psi^m\rangle=\otimes_{k=1}^{N}|\psi_{a^m_k b^m_k}\rangle$
to Bob $1$.

M5. When  Bob $j$ $(j=1,2,\cdots, n-1)$ has  received  string of $N$
qubits, he does work similarly.  First he makes eavesdropping check
using the method Alice $i$ used.   If the error rate of the samples
is higher than a threshold then he  aborts the quantum
communication. Otherwise he creates a quaternary string
$C_j=\{c_1^j, c_2^j, \cdots, c_N^j\}$ and a binary string
$D_j=\{d_1^j, d_2^j, \cdots, d_N^j\}$. Depending on the values of
$c_k^j$ and $d_k^j$ he makes operations on the qubits just  like
Alice $i$ does. That is,  he applies the   operation $\sigma_0$,
$\sigma_1$, $\sigma_2$ or $\sigma_3$
 on it depending on the corresponding   $c^j_k$ of $C_j$ is 0, 1, 2
or 3, respectively; and he implements the operation $I$ or a
Hadamard operator $H$ on the qubit according to the bit $d^j_k$ in
$D_j$ is 0 or 1, respectively.

 M6. Bob $n$ first randomly and independently chooses sufficient samples to make measurement in MB $Z$, or $X$
  randomly. Then he asks Alice 1, Alice 2, $\cdots,$ Alice $m$, and Bob 1, Bob 2, $\cdots$, Bob $n-1$
to announce publicly the $a^i_s, b^i_s, c^j_s, d^j_s$ of the samples
in a random sequential order. After that he checks the error rate of
the samples measured in the MB $Z$ or $X$ according to the XOR
results of the corresponding sample bits in the
  strings
  $B_1,B_2,\ldots,B_m, D_1, D_2, \cdots, D_{n-1}$.
  If the error rate is  higher than the
threshold $\varepsilon_t$, Bob $n$ aborts it, otherwise they go to
the next step.

 M7. All members  in group 1 and group 2 publicly announce the
strings $B_1, B_2,\ldots,B_m$ and $D_1, D_2,\ldots,D_{n-1}$. Bob $n$
 measures each of their qubits with the MB $Z$
or $X$ according to the XOR (i.e.,$\oplus$) results of the
corresponding bits in the strings $B_1,B_2,\ldots,B_m,D_1,
D_2,\ldots,D_{n-1}$.

M8. Two groups make the error rate analysis of the transmission
between the two groups. To this
 end, it  requires that all Alices and all Bobs publicly announce  $a^i_s$ and the measurement
 results of the sample qubits chosen
  randomly, and
 analyze the error rates of the samples.
  If the channel is secure, the qubit states  sent by Alice $m$  can be used as key bits for secret sharing, otherwise they discard the results obtained and retry the
 quantum communication from the beginning.

Now we  discuss the  security of this quantum secret sharing
protocol between $m$ parties and $n$ parties  with single photons
and unitary transformations. Obviously, the test in M2 and M5 can
nullify the attack with invisible photons \cite {CaiQY06} and the
Trojan horse attack \cite{DYLLZGquph05}. The checking procedure in
M5 can avoid the attack with single photons and attack with EPR
pairs \cite{YGLScienceinChina}. The fake-signal attack with EPR
pairs can be detected by the operations in M3 and the check in M2
and M5 \cite{YGLScienceinChina}. So we can believe the security of
the  protocol without question.

We should point out that the above protocol can be generalized to
the situation where three conjugate bases, or six states  are
employed \cite {GYquant062}.

In summary, we propose a  scheme for quantum secret sharing between
multi-party and multi-party with single photons and unitary
transformations, where  no
 entanglement  is employed.
In the protocol,  Alice 1 prepares a sequence of single photons in
one of four different states according to her two random classical
strings, other Alice $i$ ($2\leq i\leq m$) directly encodes her
information strings on the resulting sequence of Alice $(i-1)$ via
unitary operations, after that Alice $m$ sends the sequence of
single photons to  Bob 1.  Bob $l$ $(l=1,2,\cdots, n-1)$ does work
similar to  Alice $i$. Finally Bob $n$ measures the qubits. If
Alices
 and Bobs guarantee  the security of  quantum channel by some tests,
 then the qubit states  sent by Alice $m$  can be used as key bits for secret sharing.
  Any subset of all Alices or all
Bobs can not extract secret information, but the entire set of all
Alices and the entire set of all Bobs can. It is shown that this
scheme is in safety.
      As entanglement, especially the inaccessible multi-party entangled state, is not
 necessary in the present quantum secret sharing protocol between $m$-party and $n$-party, it may be more
 applicable when the numbers $m$ and $n$ of the parties of secret sharing are large.  This protocol is feasible with present-day technique.
\\[0.2cm]

{\noindent\bf Acknowledgments}\\[0.2cm]

 This work was supported by the National Natural Science Foundation of China under Grant No: 10671054, Hebei Natural Science Foundation of China under Grant
Nos: A2005000140, 07M006, and  the Key Project of Science and
Technology Research of Education Ministry of China under Grant No:
207011.

\end{document}